\documentclass{article}

\usepackage{arxiv}
\pdfoutput=1
\usepackage[utf8]{inputenc} 
\usepackage[T1]{fontenc}    
\usepackage{hyperref}       
\usepackage{url}            
\usepackage{booktabs}       
\usepackage{amsfonts}       
\usepackage{nicefrac}       
\usepackage{microtype}      
\usepackage{lipsum}
\usepackage[table,xcdraw]{xcolor}
\usepackage{amssymb}
\usepackage{amsmath}
\usepackage{natbib}
\usepackage{multicol}
\usepackage{multirow}
\usepackage{booktabs}
\usepackage{colortbl}
\usepackage{array}
\usepackage{enumitem}
\usepackage{bbm}
\usepackage{setspace}
\usepackage{graphicx}

\title{Transportability of Outcome Measurement Error Correction: from Validation Studies to Intervention Trials}

\author{
  Benjamin Ackerman\thanks{This is work in progress currently undergoing review. Please contact Benjamin Ackerman (backer10@jhu.edu) with any questions, comments or concerns.} \\
  Department of Biostatistics\\
  Johns Hopkins Bloomberg School of Public Health\\
  Baltimore, MD 21205 \\
  \texttt{backer10@jhu.edu} \\
   \And
 Juned Siddique\\
  Department of Preventive Medicine\\
  Northwestern University\\
  Chicago, IL 60611 \\
  \texttt{} \\
   \AND
  Elizabeth A. Stuart \\
  Departments of Mental Health, Biostatistics, and\\
  Health Policy and Management \\
  Johns Hopkins Bloomberg School of Public Health\\
  Baltimore, MD 21205
}

\begin{document}
\maketitle

\begin{abstract}
Many lifestyle intervention trials depend on collecting self-reported outcomes, like dietary intake, to assess the intervention's effectiveness. Self-reported outcome measures are subject to measurement error, which could impact treatment effect estimation. External validation studies measure both self-reported outcomes and an accompanying biomarker, and can therefore be used for measurement error correction. Most validation data, though, are only relevant for outcomes under control conditions.  Statistical methods have been developed to use external validation data to correct for outcome measurement error under control, and then conduct sensitivity analyses around the error under treatment to obtain estimates of the corrected average treatment effect. However, an assumption underlying this approach is that the measurement error structure of the outcome is the same in both the validation sample and the intervention trial, so that the error correction is transportable to the trial. This may not always be a valid assumption to make. In this paper, we propose an approach that adjusts the validation sample to better resemble the trial sample and thus leads to more transportable measurement error corrections. We also formally investigate when bias due to poor transportability may arise. Lastly, we examine the method performance using simulation, and illustrate them using PREMIER, a multi-arm lifestyle intervention trial measuring self-reported sodium intake as an outcome, and OPEN, a validation study that measures both self-reported diet and urinary biomarkers.

\end{abstract}


\doublespacing

\section{Introduction}
\label{s:intro}

Lifestyle intervention trials aim to establish how changes to human behavior, such as physical activity or food intake, can impact and improve health outcomes. In such trials, it is important to obtain accurate measures on these behaviors; however, reliable measures are often expensive and burdensome for participants to collect. Self-reported measures are therefore commonly collected as proxies of the truth. For example, food frequency questionnaires or interviewer-assisted 24-hour dietary recall may be administered in nutrition studies to quantify sodium intake instead of having participants routinely collect urine samples. While self-reported measures are more feasible to obtain, they are prone to measurement error, as participants may not be able to accurately quantify their behaviors, or may misreport their true actions \citep{willett2012nutritional}. 

Measurement error can lead to biased, less precise estimates of a treatment's effect on the outcome \citep{rothman2008modern}. In nutrition intervention studies, self-reported dietary intake measures have been shown to differ from the truth both randomly and systematically, which impacts the inferences drawn on the effectiveness of such lifestyle interventions \citep{natarajan2010measurement, espeland2001lifestyle, forster1990}. Much of the attention in the measurement error literature has been paid to when measurement error is present in either the exposure of interest or in covariates \citep{carroll2006measurement, buonaccorsi2010measurement, keogh2014toolkit}. Less work, however, has focused on correcting for misclassification or measurement error of study outcomes \citep{keogh2016statistical}, which is particularly worrisome for trials that focus on self-reported behavioral outcomes \citep{spring2018multicomponent, spring2012multiple}. This is in part because measurement error in the outcome will only lead to a biased estimate of the average treatment effect if the error is differential with respect to the intervention \citep{natarajan2010measurement}.

Existing measurement error methods, whether for covariates or outcomes, often rely on the use of a validation sample, or a group of individuals where both the truth and the observed mis-measured value are recorded \citep{wong1999measurement1, wong1999measurement2}. Such validation data can either be \emph{internal}, where the reliable biomarkers are collected on a chosen subset of individuals within the primary study, or \emph{external}, where a study is conducted on a separate set of individuals with the sole intention of modeling the error structure. Internal validation studies are typically more feasible to embed within a large observational study cohort \citep{jenab2009biomarkers}, while it is rather rare to see intervention trials, given added costs of biomarker sample collection and added burden to trial participants. External validation samples also typically only collect measures under a ``usual care" setting, which usually corresponds to a control condition in a trial, making it infeasible to directly correct for the error under both the treatment \emph{and} control conditions based on the information available to researchers. \citet{siddique2019measurement} developed methodology for modeling the measurement error under the control condition using an external validation sample, followed by sensitivity analyses to obtain a range of plausible values for the treatment effect.

While external validation samples play an important role in correcting for measurement error, concerns have been raised over external validation studies not always being ``transportable," such that the measurement error correction from an external study may not accurately apply to the main study of interest \citep{bound2001measurement, carroll2006measurement, courtemanche2015adjusting}. Previous efforts to address transportability have involved combining external validation data with internal validation data \citep{lyles2007combining}, though such an approach cannot be implemented in intervention studies without any internal validation.

In this paper, we address the issue of transportability when using external validation data to correct for measurement error of a continuous outcome in lifestyle intervention trials. Using a potential outcomes framework, we formalize cases when assumptions of measurement error transportability are violated and quantify the resulting additional bias that is introduced when estimating the average treatment effect. We then propose a weighting method for transportability, calibrating the validation data to the intervention trial to better estimate the measurement error.

\section{Definitions}
\label{s:definitions}

In order to set up the transportability issue, we will first provide some definitions and describe the measurement error problem more formally. Let $A$ denote treatment assignment (0 = control, 1 = treatment), let $S$ denote sample membership ($v$ = validation, ${rct}$ = intervention study), and let $n_s$ = the sample size of study $s$. Let $Z$ denote the outcome measured without error, $Y$ denote $Z$ measured with error, and let $X$ denote a pre-treatment covariate. $Z(a)$ and $Y(a)$ will denote potential outcomes under treatment $a$, such that $Y(a) = Z(a) + \epsilon(a)$, where $\epsilon(a) \sim N(\mu_a, \sigma_a^2)$. Here, we are assuming a simple classical measurement error structure, where the error terms are Normally distributed such that their distributions can differ across treatment groups. To expand upon this notation, let $Y^s(a)$ denote the outcome measured with error under treatment $a$ in dataset $s$. We can define the potential outcomes measured with error under different treatment conditions in different study samples in the following way:
\begin{align*}
Y^{rct}(0) &= Z(0) + \epsilon^{rct}(0) & \epsilon^{rct}(0) \sim N(\mu^{rct}_0, {\sigma^{{rct}}_0}^2) \\
Y^{rct}(1) &= Z(1) + \epsilon^{rct}(1) & \epsilon^{rct}(1) \sim N(\mu^{rct}_1, {\sigma^{{rct}}_1}^2) \\
\\
Y^v(0) &= Z(0) + \epsilon^v(0) & \epsilon^v(0) \sim N(\mu^v_0, {\sigma^{v}_0}^2) \\
Y^v(1) &= Z(1) + \epsilon^v(1) & \epsilon^v(1) \sim N(\mu^v_1, {\sigma^{v}_1}^2)
\end{align*}
Note that $Z(a)$ does not differ by study sample, because conceptually, we do not consider someone's underlying true potential outcomes to differ by which study they are in. However, we use the sample superscripts to suggest that a person's potential outcomes \emph{measured with error} can, in fact, differ by study sample, because the measurement error parameters could differ by study sample. Using these definitions, we will now formalize when the average treatment effect in the intervention trial will be biased when estimated using an outcome variable measured with error instead of a variable measured without error.

\subsection{ATE Bias under Outcome Measurement Error}
Suppose the estimand of interest in an intervention trial is the average treatment effect (ATE) of the intervention on the outcome measured without error, defined as $\Delta = E[Z(1) - Z(0)]$. However, since we do not observe $Z(0)$ or $Z(1)$ in the intervention study, we can only attempt to estimate $\Delta$ using $Y$, the observed outcome measured with error, in the intervention study with the following naive estimator:
\begin{equation}\label{eq:ATE}
\hat\Delta = \frac{\sum_{i = 1}^{n_{rct}}Y_iA_i}{\sum_{i = 1}^{n_{rct}}A_i} - \frac{\sum_{i = 1}^{n_{rct}}Y_i(1-A_i)}{\sum_{i = 1}^{n_{rct}}(1-A_i)}
\end{equation}
Using Equation \ref{eq:ATE}, we can derive the bias of $\hat\Delta$ as an estimate for $\Delta$ as:
\begin{equation}\label{eq:ATEbias}
\text{bias}_{\hat\Delta} = \mu^{rct}_{1} - \mu^{rct}_{0}
\end{equation}
In other words, estimating $\Delta$ using the mis-measured outcome will be a biased estimate if the means of the outcome error under treatment and control conditions are different. Note that if the measurement error is either classical or differential with respect to treatment, but the errors under treatment and control conditions are centered around the same value, then $\hat\Delta$ will still be an unbiased estimate of $\Delta$ (its variance may be inflated, though this is not the focus of the current paper). Throughout this paper, we will therefore focus on the case where the measurement error is differential with respect to treatment, such that in the trial, $\mu^{rct}_{0} \neq \mu^{rct}_{1}$. 

One challenge to correcting for $\text{bias}_{\hat\Delta}$ is that the true outcomes measured without error, $Z(0)$ and $Z(1)$, are typically unobserved in an intervention trial. Therefore, the error means $\mu^{rct}_{0}$ or $\mu^{rct}_{1}$ cannot be estimated using the trial data alone. One strategy is to utilize information from an external validation study to estimate $\mu^{rct}_{0}$ or $\mu^{rct}_{1}$ by making an assumption of transportability. However, external validation studies typically only measure the outcomes under a single control condition, as described in Table \ref{tab:observables} below.

\begin{table}[h]
\centering
\caption{Observables by study sample. Cells shaded in grey denote observed measures, while blank cells denote unobserved quantities.}\label{tab:observables}
\begin{tabular}{|rl|c|c|c|c|c|}
\hline
                   &       & Z(0)                     & Z(1) & Y(0)                                            & Y(1)        &  X                                  \\ \hline
validation         & S = v & \cellcolor[HTML]{C0C0C0}$\checkmark$ &      & \cellcolor[HTML]{C0C0C0}  $\checkmark$                      &          &     \cellcolor[HTML]{C0C0C0}  $\checkmark$                                 \\ \hline
intervention study & S = ${rct}$ &                          &      & \cellcolor[HTML]{C0C0C0}{\color[HTML]{000000} } $\checkmark$ & \cellcolor[HTML]{C0C0C0}{\color[HTML]{000000} } $\checkmark$&\cellcolor[HTML]{C0C0C0}{\color[HTML]{000000} }$\checkmark$\\ \hline
\end{tabular}
\end{table}

Note that if the potential outcomes under treatment, $Z(1)$ and $Y(1)$, were also observed in the validation sample, then the validation study would be considered an intervention trial in itself. Such a scenario is highly unlikely, given the intention and design of external validation studies. To account for this, \citet{siddique2019measurement} propose using external validation data to estimate the error mean under control conditions, $\mu_0^{rct}$. We will consider the following naive estimator for $\mu_0^{rct}$:
\begin{equation}\label{eq:mu0hatnaive}
\hat\mu_0^{naive} = \frac{1}{n_v}\sum_{i = 1}^{n_v}(Y_i - Z_i)
\end{equation}
Observe that $\hat\mu_0^{naive}$ is an unbiased estimate of $\mu_0^v$, the error mean under control in the \emph{validation sample}. By assuming that transportability holds, we are also assuming that it is an unbiased estimate for $\mu_0^{rct}$, the error mean under control in the \emph{intervention study}.

Lastly, \citet{siddique2019measurement} conduct sensitivity analyses around the error under treatment ($\mu_1^{rct}$) to obtain a plausible range of estimates for $\Delta$. We build on this work by proposing a solution to when the transportability assumption is violated, and $\hat\mu_0^{naive}$ is therefore a biased estimate for the error mean under control in the intervention study, $\mu_0^{rct}$.

The remainder of the paper is structured as follows: in Section \ref{s:transportability}, we describe the transportability assumption evoked to estimate $\mu_0^{rct}$ using validation data and discuss its plausibility. We then formalize when the transportability assumption will be violated, and how much bias is introduced as a result. Then, in Section \ref{s:weighting}, we propose the utilization of propensity score-type weighting methods to decrease the bias of estimating $\mu_{0}^{rct}$ using validation data, followed by a simulation illustrating the performance of the weighting methods in Section \ref{s:simulation}. We apply the methods to a data example in Section \ref{s:example}, and conclude with a discussion of outcome measurement error and the utilization of validation data, along with some limitations, in Section \ref{s:discuss}.

\section{Transportability}
\label{s:transportability}

Suppose $\hat\Delta$ is a biased estimate of $\Delta$ in an intervention trial, and external validation data is therefore utilized to partially account for $\text{bias}_{\hat\Delta}$ by estimating the mean error under control, $\mu_0^{rct}$. In order to obtain an estimate for $\mu_0^{rct}$, a transportability assumption must be made. Formally, the assumption is as follows:
$$f(Y^{v}(1),Y^{v}(0)|Z(1),Z(0),X) = f(Y^{rct}(1),Y^{rct}(0)|Z(1),Z(0),X)$$
This transportability assumption implies that, under Normality, $\mu^v_{0} = \mu^{rct}_{0}$ and ${\sigma^v_{0}}^2 = {\sigma^{rct}_{0}}^2$ (and also that $\mu^v_{1} = \mu^{rct}_{1}$ and ${\sigma^v_{1}}^2 = {\sigma^{rct}_{1}}^2$). In other words, the assumption states that the measurement error structures for the potential outcomes in the validation sample are the same as they are in the intervention study. This assumption also implies that $\hat\mu_0^{naive}$, which estimates the error mean under control using validation data, is an unbiased estimate of $\mu^{rct}_{0}$. However, the transportability assumption may not hold in some cases, which would introduce additional bias when estimating the $\Delta$. We will now describe when the transportability assumption will be violated.

\subsection{Bias from transportability violation ($\text{bias}_{\hat\mu_0}$)}

In order to formalize when there will be bias in estimating $\mu^{rct}_{0}$ using validation data, consider the case where we have a single covariate, $X$, such that $X = \beta_0 + \beta_1 \mathbbm 1_{S=v} + \epsilon_X$, where $\epsilon_X$ has mean 0 and variance $\sigma_X^2$. Observe that $\beta_0 = E[X|S = {rct}]$ and $\beta_1 = E[X|S=v] - E[X|S={rct}]$. In other words, $\beta_1$ represents the difference in mean of covariate $X$ across the two datasets (trial and validation data).

Next, recalling the classical measurement error structure of $Y = Z + \epsilon$, consider when the error term is distributed as $\epsilon \sim N(\alpha_0 + \alpha_1A + \alpha_2X, \sigma_Y^2)$ such that the measurement error is differential with respect to both treatment and $X$. By performing a substitution for $X$, we obtain the following:

\begin{equation}\label{eq:epsilon}
\epsilon \sim N(\alpha_0 + \alpha_1A + \alpha_2\{\beta_0 + \beta_1 \mathbbm1_{S = v} + \epsilon_X \}, \sigma_Y^2)
\end{equation}

Recall that $\mu^s_{a} = E[Y^s(a)] - E[Z(a)]$. The measurement error mean parameters under each treatment condition in each dataset can therefore be expressed as follows:
\begin{align*} 
  \mu^{rct}_{0} &= \alpha_0 + \alpha_2 \beta_0 \\
  \mu^{rct}_{1} &= \alpha_0 + \alpha_2 \beta_0 + \alpha_1  \\
  \mu^v_{0} &= \alpha_0 + \alpha_2 (\beta_0 + \beta_1) \\
  \mu^v_{1} &= \alpha_0 + \alpha_2 (\beta_0 + \beta_1) + \alpha_1 
\end{align*}
First, notice that $\text{bias}_{\hat\Delta}$, which is the difference in error means between treatment and control conditions in the trial, can be expressed as $\alpha_1$. Next, we can derive the bias of $\hat\mu_0^{naive}$, which uses validation data, as an estimate of $\mu_0^{rct}$ as follows: 
\begin{equation}\label{eq:biasmu0}
\text{bias}_{\hat\mu_{0}} = \mu^v_{0} - \mu^{rct}_{0} = \alpha_2 \beta_1
\end{equation}
The transportability assumption will therefore be violated if $\alpha_2 \neq 0$ \emph{and} $\beta_1 \neq 0$. In other words, if a covariate $X$ impacts the measurement error structure ($\alpha_2 \neq 0$), \emph{and} the distribution of $X$ differs across the trial and the validation sample ($\beta_1 \neq 0$), then $\hat\mu_0^{naive}$ will be a biased estimate of $\mu_0^{rct}$. This also extends to when there are multiple covariates that meet these two conditions. 

The transportability assumption violation, and the introduction of $\text{bias}_{\hat\mu_0}$, may also increase $\text{bias}_{\hat\Delta}$. Observe that if we substitute the estimate $\hat\mu_0^{naive}$ for $\mu_0^{rct}$ in Equation \ref{eq:ATEbias}, we obtain the following:
\begin{align*}
\widetilde{\text{bias}}_{\hat\Delta} &= \mu_1^{rct} - \hat\mu_0^{naive}\\
\mathbb E[\widetilde{\text{bias}}_{\hat\Delta}] &= \mu_1^{rct} - \mu_0^{rct} - \alpha_2\beta_1 \\
&= \alpha_1 - \alpha_2\beta_1
\end{align*}
This motivates the proposal of a weighted estimator for $\mu_0^{rct}$, that reduces $\text{bias}_{\hat\mu_0}$, which we present in Section \ref{s:weighting}.

Table \ref{tab:biascases} summarizes the discussion on the two potential biases, $\text{bias}_{\hat\Delta}$ and $\text{bias}_{\hat\mu_0}$, providing different cases for researchers to consider when these biases should be of concern. While this may not be an exhaustive list of \emph{all} possible scenarios, we think of these as the most plausible scenarios in practice that researchers may encounter. Only Scenario VI, in which the measurement error is differential in the trial with respect to $A$, differential in the trial and validation sample with respect to $X$, and the distribution of $X$ differs between the trial and the validation sample, violates the transportability assumption. Scenario V is technically possible, where $\text{bias}_{\hat\mu_0} \neq 0$ while $\text{bias}_{\hat\Delta} = 0$. However, note that the motivation for outcome measurement error correction is only really when the measurement error is differential by treatment (i.e. $\text{bias}_{\hat\Delta} \neq 0$), so Scenario V is therefore highly unlikely.

\begin{table}[]
\caption{Conditions under which to be concerned about bias in the $ATE$ ($\text{bias}_{\hat\Delta}$) and/or bias in the measurement error (ME) correction ($\text{bias}_{\hat\mu_0}$)}
\label{tab:biascases}
\centering \begin{tabular}{|>{\centering\arraybackslash}c|p{2cm}|p{2cm}|p{2cm}|p{2cm}|p{2cm}|}
\hline
Scenario & ME differs \newline by A & ME differs \newline by X & X differs\newline by sample & $\text{bias}_{\hat\Delta}$ & $\text{bias}_{\hat\mu_0}$ \\
\hline
I        &                                  &                          &                                                     & 0 \newline \newline $\mu_0^{rct} = \mu_1^{rct}$                   & 0                     \\
\hline
II        &  \centering{$\checkmark$}                         &                          &                                                     & $\alpha_1$ \newline \newline $\mu_0^{rct} \neq \mu_1^{rct}$            &    0             \\
\hline
III        &  \centering{$\checkmark$}                         &  \centering{$\checkmark$}                &                                                     & $\alpha_1$    \newline \newline $\mu_0^{rct} \neq \mu_1^{rct}$       &  0 \newline \newline $\alpha_2 \neq 0$, \newline but $\beta_1 = 0$                    \\
\hline
IV        &  \centering{$\checkmark$}                         &                          & \centering{$\checkmark$}                                           & $\alpha_1$  \newline \newline $\mu_0^{rct} \neq \mu_1^{rct}$         & 0 \newline \newline $\beta_1 \neq 0$, \newline but $\alpha_2 = 0$                     \\
\hline
V        &                         &  \centering{$\checkmark$}                 & \centering{$\checkmark$}               & 0 \newline \newline $\mu_0^{rct} = \mu_1^{rct}$          &  $\alpha_2 \beta_1$ \newline \newline $\beta_1 \neq 0$ \newline and $\alpha_2 \neq 0$ \\
\hline
VI        &  \centering{$\checkmark$}                         &  \centering{$\checkmark$}                 & \centering{$\checkmark$}                             & $\alpha_1$  \newline \newline $\mu_0^{rct} \neq \mu_1^{rct}$          &  $\alpha_2 \beta_1$ \newline \newline $\beta_1 \neq 0$ \newline and $\alpha_2 \neq 0$ \\
\hline
\end{tabular}
\end{table}

\subsection{Additional Assumptions}

In addition to the transportability assumption, we make a parametric assumption that the measurement error model form is the same across the two samples. In other words, we assume that if the measurement error is differential with respect to a given covariate in the validation sample, then it is also differential with respect to that covariate in the intervention trial (i.e. if age impacts the measurement error structure in the validation sample, it also does so in the trial). This assumption extends to the presence of higher order terms, such as interactions or quadratic terms, that they be present in both samples. We also must assume that there are no unobserved covariates that impact the measurement error and differ between the two samples. Lastly, we must make an assumption of common support, that the range of all covariates in the validation sample are covered by their respective ranges in the intervention trial. For example, we cannot transport an estimate from a validation study where the oldest participant is fifty years old to an intervention trial with participants over the age of fifty. Another way to frame this is that each trial participant has a nonzero probability of participating in the external validation study. The plausibility of these assumptions are discussed further in Section \ref{s:discuss}.

\section{Weighting-Based Approach to Reduce Transportability Bias}
\label{s:weighting}

We will now describe the use of propensity score-type weights to reduce the transportability bias. Propensity scores have been traditionally used in non-experimental studies, where treatment is not randomized, to make treatment groups more similar on pre-treatment covariates using matching or weighting methods \citep{rosenbaum1983central}. This approach has since been applied to the fields of transportability and generalizability, where propensity scores are used to model the conditional probability of trial participation (instead of treatment assignment). The probabilities are subsequently used to weight a randomized trial so it better resembles a well-defined target population on observed pre-treatment covariates \citep{stuart2018generalizability, kern2016assessing, dahabreh2018generalizing}.

Previous work has demonstrated similar benefits of implementing propensity score-type weighting methods when using external validation data to adjust for missing confounders \citep{mccandless2012adjustment} and when evaluating disease prediction models in samples that differ from the target population \citep{powers2019evaluating}. Here, we are interested in addressing the transportability violation by weighting the external validation sample so that it better resembles the intervention study of interest on a set of observed pre-treatment covariates.

In brief, we will do so by modeling the probability of study membership (trial vs. validation study), and then weighting the validation sample before estimating $\mu_0$. Consider the following model of study participation:

\begin{equation}\label{psmodel}
\text{logit}(P(S = rct)) = \theta^t X
\end{equation}

Where $X$ is a set of observed covariates measured in both the trial and the validation data. We can then predict the probability of trial participation as 
\begin{equation}\label{trialmembership}
\hat e_i = \hat e(X) = \text{expit}(\hat \theta^t X)
\end{equation}
Similar to ATT weighting for non-experimental studies, we then construct the following weights:
\begin{equation}\label{weights}
\hat w_i = \begin{cases} \frac{\hat e_i}{1- \hat e_i} & \text{ if } S = v\\
0 & \text{ if } S = {rct}
\end{cases}
\end{equation}
Using these weights, we can then estimate $\mu_0^{rct}$ using validation data with the following estimator:
\begin{equation}\label{eq:mu0hatweighted}
\hat\mu_0^{weighted} = \frac{\sum_{i=1}^{n_v} \hat w_i(Y_i-Z_i)}{\sum_{i=1}^{n_v}\hat w_i}
\end{equation}
Individuals in the validation sample that are more similar to the trial participants will have greater predicted probabilities of being trial participants, and will therefore have larger weights. Members of the external validation sample that are most dissimilar to the trial sample will be down-weighted towards zero. In this way, we can obtain a weighted estimate of the error mean under control in the validation sample, $\mu_0^{v}$, such that we reduce the bias of this estimate as an estimate for $\mu_0^{rct}$ due to covariate differences across samples. Details on estimating the standard error for inference can be found in the Supplemental Materials.

\subsection{Weighting under Misspecification of the Sample Membership Model}

Recall that the error estimate in the validation sample will be a biased estimate of the error in the trial if there exists a set of covariates that impact the measurement error structure and that also differ by sample. We therefore want to include all observed covariates that fall into this category when fitting the model of sample membership. If we fit the \emph{correct} model of sample membership accounting for all such Xs in the true form, then we should be able to eliminate the bias of our $\mu_0$ estimate through this weighting procedure (as is the case when transporting trial results to a target population using inverse odds weighting, see proof in \citet{westreich2017transportability}). In practice, however, it can be quite challenging to fit the correct sample membership model (i.e. there may be complex interactions or higher order terms in the true model that are omitted). Fitting a simpler, misspecified sample membership model may lead to a smaller reduction of the bias when weighting. In the next section, we describe a simulation study, where we demonstrate the performance of the proposed weighting methods under increasingly complex true sample membership models, and varying amounts of model misspecification. 

\section{Simulation}
\label{s:simulation}

We now conduct a simulation study to assess the weighting methods described in Section \ref{s:weighting} on decreasing $\text{bias}_{\hat\mu_0}$. We consider four covariates, and vary the following: (1) the degree to which each $X$ impacts the measurement error model, (2) the degree to which each $X$ impacts the trial membership model, and (3) the degree of misspecification of the trial membership model that is fit using the validation sample.

\subsection{Simulation Setup}

Consider the following measurement error model of $Y$:

$$Y = \alpha_0 + \alpha_1 Z + \alpha_2 A + \alpha_3 X_1 + \alpha_4 X_2 + \alpha_5 X_3 + \alpha_6 X_4 + \epsilon_Y $$

where $\epsilon_Y \sim N(0, \sigma_Y^2)$. Since we only require that the $X$s impact the error structure in \emph{some} capacity, we do not consider other, more complex true structures of the measurement error model in this simulation study.

We vary the true underlying models of sample membership by considering the following seven model forms: 

\begin{enumerate}
    \item $S \sim X_1 + X_2 + X_3 + X_4$
    \item $S \sim X_1 + X_2 + X_3 + X_4 + X_3^2$
    \item $S \sim X_1 + X_2 + X_3 + X_4 + X_4^2$
    \item $S \sim X_1 + X_2 + X_3 + X_4 + X_3 X_4$
    \item $S \sim X_1 + X_2 + X_3 + X_4 + X_3 X_4 + X_3^2 + X_4^2$
    \item $S \sim X_1 + X_2 + X_3 + X_4 + X_1X_4$
    \item $S \sim X_1 + X_2 + X_3 + X_4 + X_1X_3$
\end{enumerate} 

We parameterize the coefficients for the covariates $X_1$, $X_2$, $X_3$, and $X_4$ as $\{\gamma_1, 0, \frac{1}{2}\gamma_1, 2\gamma_1\}$ in the measurement model, and as $\{0, \gamma_2, 2\gamma_2, \frac{1}{2}\gamma_2\}$ in the trial membership model (See Supplemental Table \ref{tab:simparams}). In doing so, we establish that covariate $X_1$ impacts sample membership but not measurement error, $X_2$ impacts measurement error but not sample membership, $X_3$ weakly impacts sample membership and strongly impacts measurement error, and $X_4$ strongly impacts sample membership and weakly impacts measurement error. We set any quadratic term coefficients to $50\%$ of the original X's coefficient (i.e. if $X_3$ has a coefficient of $\frac{1}{2}\gamma_1$, then $X_3^2$ would have a coefficient of $\frac{1}{4}\gamma_1$). For interaction terms, the coefficient is set to the average of the two Xs' original coefficients (i.e. the coefficient for a $X_1 X_3$ interaction term would be $\frac{3}{4}\gamma_1$). 

The two parameters $\gamma_1$ and $\gamma_2$ function as scaling parameters, varying from 0 to 1 by increments of 0.2. Observe that when $\gamma_1 = 0$, the four covariates do not impact sample membership at all, and the trial membership probabilities are expected to be $0.5$ in both study samples. As $\gamma_1$ increases to 1, the impact of the variables on sample membership increases, and the overlap of the probabilities across the two samples decreases. In this way, $\gamma_1$ can be considered a function of the absolute standardized mean difference (ASMD) of the participation probabilities between the trial and the validation sample. When $\gamma_2 = 0$, then the measurement error is not differential with respect to any of the covariates, and as $\gamma_2$ increases, so does the impact of the covariates on the measurement error structure. Note that when either $\gamma_1 = 0$ or $\gamma_2 = 0$, then we expect that $\text{bias}_{\hat\mu_0} = 0$.

For each of the seven true trial membership models, we fit both the true model, which is correctly specified, as well as a main-effects-only model. The main-effects-only model will be misspecified when the true model has interaction and/or quadratic terms. This type of misspecification illustrates a plausible scenario, in which researchers may fit a simple multi-variable logistic regression model to estimate the trial participation probabilities, ignoring potential complexities in the underlying true model form.

In order to quantify the degree of model misspecification (DoM), \citet{lenis2018measuring} propose a unit-independent, informative metric:

$$\eta_S = \frac{1}{N}\sum^{N}_{i=1} \frac{|\hat{\pi_i} - \hat {\pi_i^C}|}{\sigma_{\hat \pi^C}}$$
where $\hat \pi_i$ is the predicted probability of being in the trial under the specified model, and $\hat \pi_i^C$ is the predicted probability of being in the trial under the true selection model. We use the DoM metric to relate the amount of model misspecification across scenarios.

In total, there are 756 simulation scenarios that vary by: (6 values for $\gamma_1$) $\times$ (6 values for $\gamma_2$) $\times$ (7 true trial membership models) $\times$ (3 weighting options: unweighted naive estimator, weighted estimator with correctly specified weights, and weighted estimator with misspecified weights). We iterate over each scenario 1000 times, and will now describe the data generation process.

\subsection{Data Generation}
Consider one particular scenario from the 756 scenarios outlined above. We start by simulating a population of four $X$ covariates ($N = 1000000$) according to a multivariate Normal distribution with mean 0, variance 1, where there is no correlation between the $X$s. Based on the scenario's $\gamma_1$ value and true trial membership model form - suppose the simplest true form for example - we generate the probability of being in the trial (vs. the validation sample) for the whole population as follows: 
$$p_i = \text{expit}(\gamma_1X_{1i} + 0X_{2i} + \frac{1}{2}\gamma_1 X_{3i} + 2\gamma_1X_{4i})$$
for $i = 1,...,N$. Next, we generate the binary sample membership variable $S$ for each member of the population as $S_i \sim \text{Bernoulli}(p_i)$ for $i = 1,...,N$. Note that while each $p_i$ is determined by the scenario's specified parameters, $S$ is assigned with a degree of randomness, such that each person in the population theoretically has a chance of being ``in the trial" or ``in the validation sample" across each different simulation run.

After assigning $S$, we randomly sample members for the trial and validation samples, each of size $n = 1000$. In this step, we observe the differences in the covariates across the two samples as specified by the $\gamma_1$ scaling parameter. We then generate the true potential outcomes $Z(0)$ and $Z(1)$ as $Z(0) \sim N(0, 1)$ and $Z(1) \sim N(2, 1)$, and the mis-measured potential outcomes $Y(0)$ and $Y(1)$ as:
$$Y(a) \sim N\big( Z(a) + 0X_1 + \gamma_2X_2 + 2\gamma_2X_3 + \frac{1}{2}\gamma_2X_4, 1.5 \big )$$ such that the variance of $Y(a)$ is $1.5$ times the variance of $Z (a)$.

We assign treatment $A$ as Bernoulli$(0.5)$ in the trial and 0 in the validation sample. Lastly, we generate the observed outcomes $Z$ and $Y$ as $Z = A\times Z(1) + (1-A)\times Z(0)$ and $Y = A\times Y(1) + (1-A)\times Y(0)$.

\subsection{Simulation Results}

\begin{figure}
 \centerline{\includegraphics[width=145mm]{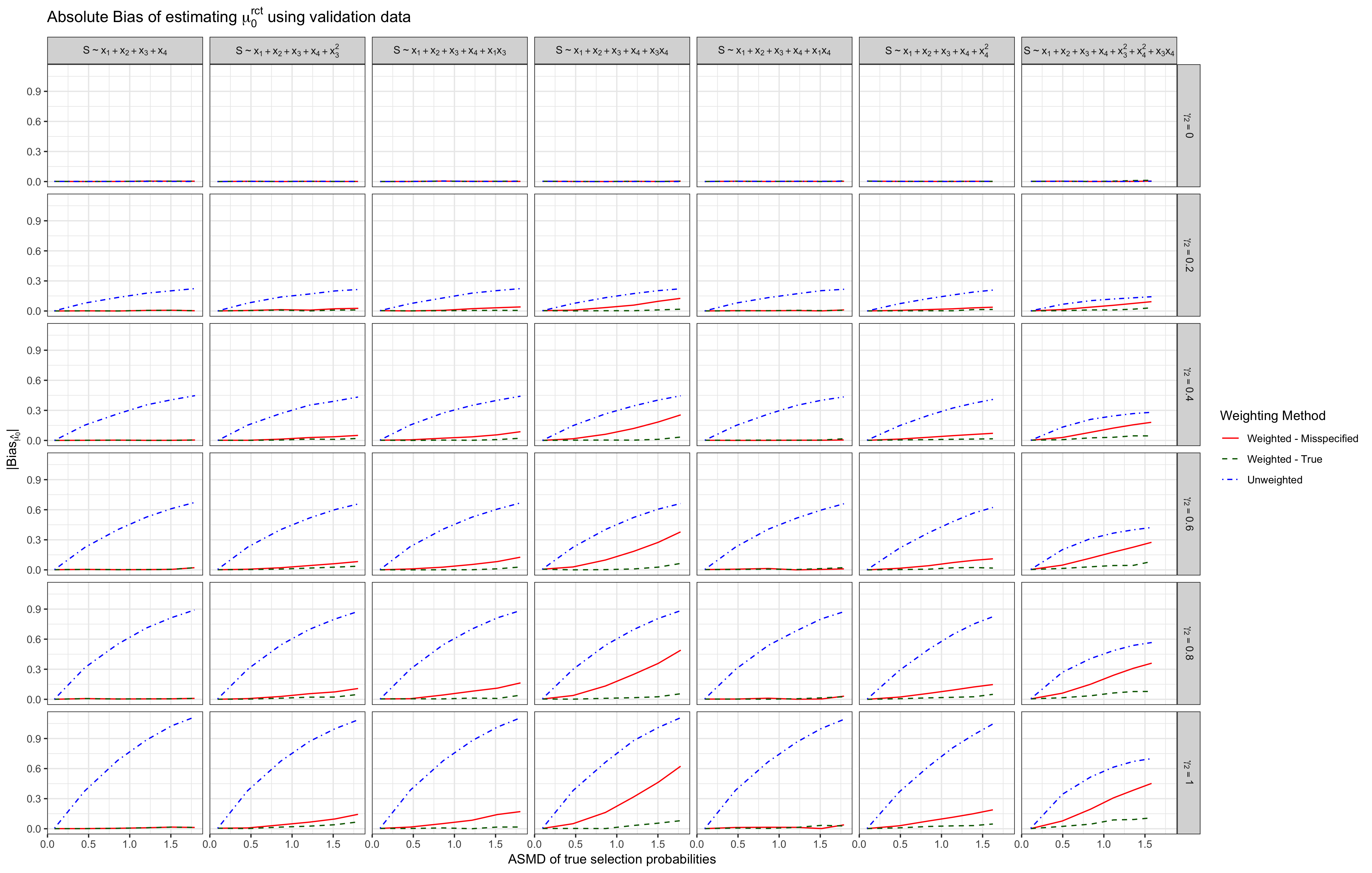}}
\caption[Bias of estimating the error mean under control using validation data]%
{Bias of estimating the error mean under control using validation data. Each column represents a different true sample membership model. From top to bottom row, the $\gamma_2$ ``scale" parameter for the impact of the Xs on the measurement error increases, meaning the strength of the relationship between Y and the Xs is increasing in magnitude. The different line types and colors represent the different weighting approaches: Unweighted (blue dotted dash), weighted by fitting the simplest additive model (``Weighted - Misspecified", red solid), and weighted by fitting the true selection model (``Weighted - True", green dash). This figure appears in color in the electronic version of this article.}
\label{fig:simresults}
\end{figure}

Figure \ref{fig:simresults} shows the absolute $\text{bias}_{\hat\mu_0}$, or the transportability bias, across simulation scenarios. Each column represents a different underlying true sample membership model structure, increasing in complexity and degree of misspecification from left to right (see Supplemental Figure \ref{fig:dommodel} for DoM by sample membership model form). The strength of the impact of the $X$s on the measurement error ($\gamma_2$) increases by row, from top to bottom. The x axis depicts the absolute standardized mean difference (ASMD) of the true selection probabilities across the samples, representing an increasing difference between the $X$ distributions across the samples ($\gamma_1$) (see Supplemental Figure \ref{fig:asmdscale}). Note that scenarios where the ASMD is greater than 1 represent fairly extreme, less realistic settings. The three lines represent the absolute bias of the naive estimator, the weighted estimator with misspecified weights, and the weighted estimator with correctly specified weights.

\begin{figure}
 \centerline{\includegraphics[width=145mm]{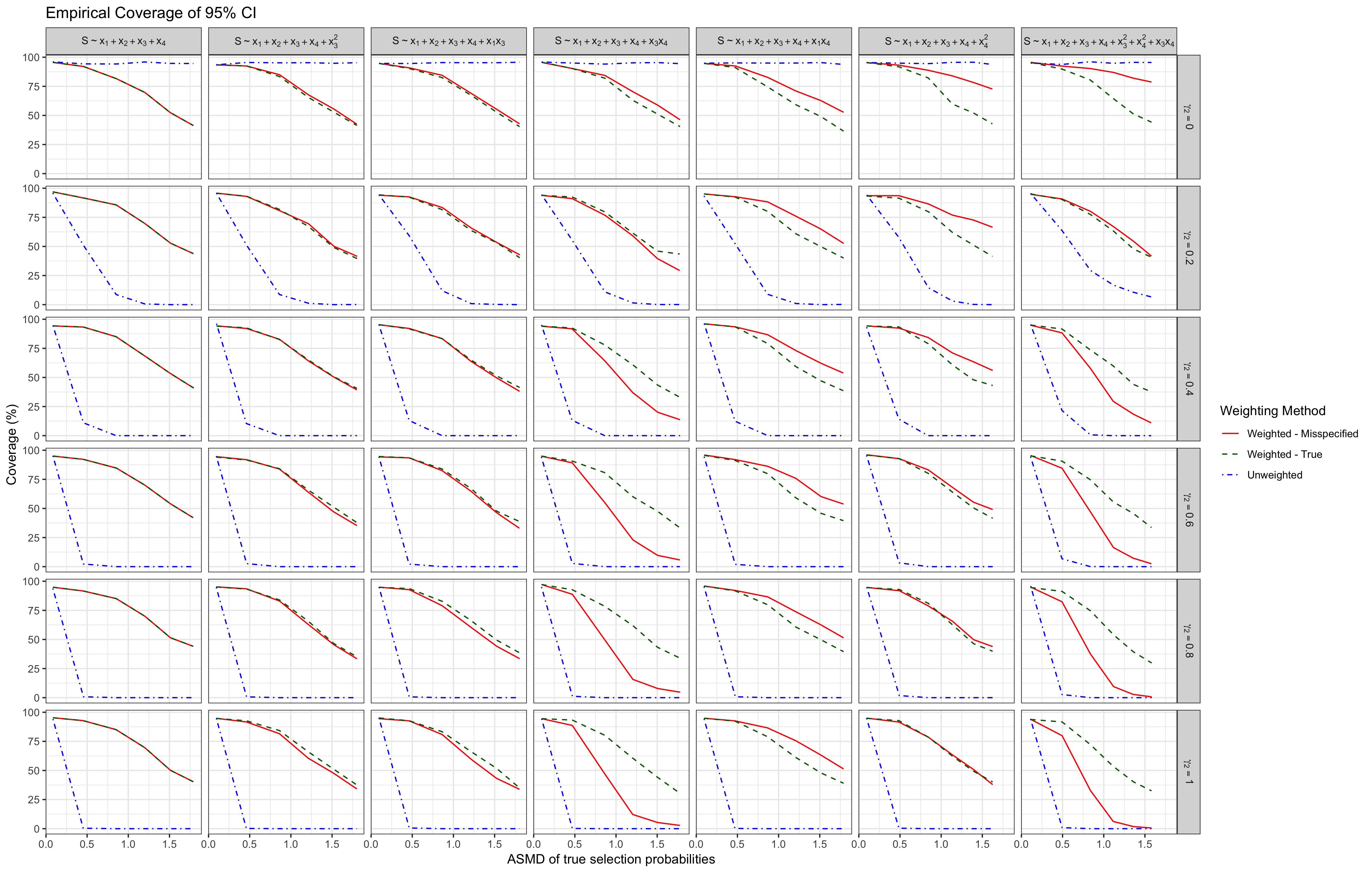}}
\caption[Empirical coverage of 95\% CI for estimators of $\mu_0^{rct}$]%
{Empirical coverage of 95\% CI for estimators of $\mu_0^{rct}$. Each column represents a different true sample membership model. From top to bottom row, the $\gamma_2$ ``scale" parameter for the impact of the Xs on the measurement error increases, meaning the strength of the relationship between Y and the Xs is increasing in magnitude. The different line types and colors represent the different weighting approaches: Unweighted (blue dotted dash), weighted by fitting the simplest additive model (``Weighted - Misspecified", red solid), and weighted by fitting the true selection model (``Weighted - True", green dash). This figure appears in color in the electronic version of this article.}
\label{fig:simresults_coverage}
\end{figure}

First, note that in the top row, the absolute bias is zero under all cases, because the measurement error is not differential with respect to any of the covariates (Table \ref{tab:biascases}, Scenarios II and IV). Next, note that in all of the plots, the absolute bias is zero when the ASMD is zero, or when the distribution of covariates in the trial and validation sample do not differ (Table \ref{tab:biascases}, Scenarios II and III). Under the same true sample membership model (for a given column), as the impact of the $X$s on the measurement error model increases (from top to bottom row), and as the distributional difference of the $X$s increases between the samples (from left to right of the x axis), the absolute $\text{bias}_{\hat\mu_0}$ also increases.

When the selection model is correctly specified, and the resulting predicted probabilities are used to construct the weights, the weighted estimator is nearly unbiased in all scenarios (except for when the impacts of the $X$s on the measurement error model and on sample membership are extremely, somewhat unrealistically, large). In practice, though, model misspecification is very plausible, as it may be common for researchers to fit a main-effects-only model of the covariates to predict the sample membership probabilities, unable to identify the true model form. Under varying amounts of model misspecification, we see that the weighted estimator still performs fairly well in reducing $\text{bias}_{\hat\mu_0}$. As the severity of the transportability assumption violation increases, the weighted estimator with misspecified weights does appear to perform worse than the weighted estimator with correctly specified weights under certain scenarios, particularly when the omitted interaction and/or quadratic terms are for covariates that more strongly impact sample membership and measurement error. For example, in column 4, the main-effects-only model omits an interaction between $X_3$ and $X_4$, the two variables that impact \emph{both} models. That weighted estimator performs worse than the main-effects-only model that omits an $X_1 X_3$ interaction (column 3), and the model that omits an $X_1 X_4$ interaction (column 4), as $X_1$ does not impact the measurement error structure at all.

Figure \ref{fig:simresults_coverage} shows the empirical 95\% confidence interval coverage of the different estimators for $\mu_0^{rct}$. First, observe that across all scenarios in which the covariates impact the measurement error (rows 2-6), the coverage of the naive estimator sharply decreases towards zero as the intervention trial differs more greatly from the validation sample on pre-treatment covariates. Note also, though, that the weighting approaches (even with a misspecified model) generally yield substantially better confidence interval coverage than does the naive approach. This aligns with the pattern in Figure \ref{fig:simresults}, in which the naive estimator becomes increasingly biased as the samples become less similar on the covariates. Next, note that the weighted estimator with correctly specified weights tends to have better coverage than the weighted estimator with misspecified weights, though the coverage also decreases in the more extreme cases of covariate differences between the samples. Still, the rate of coverage decline for the weighted estimators by covariate differences is far smaller than that of the naive estimator. Also, note that for the more plausible scenarios (ASMD < 1), the coverage of the weighted estimators is still fairly good, and far better than that of the naive estimator.

In the cases where the measurement error is \emph{not} differential with respect to the Xs (top row), the weighted estimators seem to have worse coverage than the naive estimator, even though they are all unbiased. 
This is due to the presence of large/extreme weights, which shift the point estimates of $\mu_0^{rct}$ further away from the truth and therefore increase the variability of the weighted estimators, even though they are still unbiased. Trimming extreme weights (especially in the less plausible scenarios where ASMD > 1), resulted in better coverage across these settings. However, in cases where the measurement error is not differential with respect to covariates, and the transportability assumption is not believed to be violated, then the weighting approach may not be preferable.

Overall, though, it appears that \emph{any} weighting, whether by fitting a main-effects-only model, the true selection model, or anything in between, greatly improves the transportability of the control group error mean from the validation sample to the trial. When there are concerns about measurement error corrections not transporting properly from an external validation sample to an intervention trial, these simulation results highlight that the weighting method proposed in Section \ref{s:weighting} can help reduce the bias and improve coverage due to poor transportability.

\section{Data Example}
\label{s:example}

We now apply the methods described in Section \ref{s:weighting} to a lifestyle intervention trial, PREMIER, using OPEN, an external validation sample.

In PREMIER, 810 individuals were randomized to either one of two behavioral/dietary recommendations, or to standard care, to estimate the effect of the intervention on blood pressure reduction \citep{svetkey2003premier}. For illustrative purposes, instead of blood pressure, we focus on self-reported sodium intake, measured by 24-hour recall, as the outcome of interest. We also combine the two intervention groups into one ``treatment" group. PREMIER is a rather unique intervention trial, in that urinary sodium intake was \emph{also} collected at each time point in addition to the self-reported intake, providing an opportunity for us to evaluate the method performance. Note that this is atypical for an intervention trial to collect. We use sodium intake at 18 months followup as the outcome of interest (though note that in the original trial, the primary time point of interest for analysis was 6 months), and we limit the sample only to those who have both self-reported and urinary sodium intake measures at 18 months (n = 670).

OPEN is an external validation study that measures both self-reported sodium (via 24-hour recall) and urinary sodium (via a 24-hour urine sample) on a sample of 484 study participants, with the goal of understanding the structure and amount of measurement error among self-reported dietary outcomes \citep{subar2003using}. Using PREMIER and OPEN, we will demonstrate that the measurement error of dietary sodium intake is differential with respect to pre-treatment covariates, and that the distribution of these factors also differ between the intervention trial and the validation study. Therefore, the transportability assumption is violated, and $\hat\mu_0^{naive}$, which is estimated in OPEN, is a biased estimate for $\mu_0^{rct}$ in PREMIER.

Table \ref{tab:covtab} describes the distribution of covariates across the two studies, along with the ASMD of each covariate between the two studies. Observe that BMI differs greatly between the two studies. Additionally, the OPEN population appears to be older, more male and less racially diverse than PREMIER.

\begin{table}

\caption{\label{tab:}Distribution of Covariates by Study \label{tab:covtab}}
\centering
\begin{tabular}[t]{lrrr}
\toprule
  & OPEN (n=484) & PREMIER (n=810) & ASMD\\
\midrule
\rowcolor{gray!6}  Male & 0.54 & 0.38 & 0.33\\
\addlinespace[0.3em]
\multicolumn{4}{l}{\textbf{Age}}\\
\hspace{1em}$\leq 40$ & 0.03 & 0.14 & 0.42\\
\rowcolor{gray!6}  \hspace{1em}41-45 & 0.17 & 0.17 & 0.01\\
\hspace{1em}46-50 & 0.21 & 0.24 & 0.06\\
\rowcolor{gray!6}  \hspace{1em}51-55 & 0.20 & 0.21 & 0.03\\
\hspace{1em}56-60 & 0.15 & 0.13 & 0.05\\
\rowcolor{gray!6}  \hspace{1em}$\geq$ 61 & 0.24 & 0.12 & 0.30\\
BMI & 27.87 & 33.06 & 0.95\\
\rowcolor{gray!6}  Black & 0.06 & 0.34 & 0.81\\
\addlinespace[0.3em]
\multicolumn{4}{l}{\textbf{Education}}\\
\hspace{1em}College & 0.55 & 0.59 & 0.08\\
\rowcolor{gray!6}  \hspace{1em}Grad School & 0.32 & 0.32 & 0.00\\
\bottomrule
\end{tabular}
\end{table}

In order to implement the methods described in Section \ref{s:weighting}, we form a ``stacked" dataset, comprised of data from PREMIER and OPEN, which contains variables for sample membership ($S$), treatment ($A$), self-reported dietary sodium intake ($Y$), urinary sodium intake ($Z$), and the following common covariates ($X$): age category, sex, race, BMI and education. Certain covariates, like age and education, are categorized to ensure consistency in measures across datasets, and race is utilized as a dichotomous variable, indicating if individuals identify as Black or not. Again, note that typically $Z$ would be coded as missing when $S = rct$; however, the unique nature of PREMIER allows us to observe Z in the trial.

By comparing the difference in outcome means by measurement type in PREMIER, it appears that the self-reported dietary sodium measures under-report the true sodium intake in both treatment arms, and $\text{bias}_{\hat\Delta}$ at 18 months is estimated to be $0.028$ (see Supplemental Table \ref{tab:outcomes}). While this difference is not significant (see Supplemental Table \ref{tab:errortx}), we still proceed to assess the feasibility of transporting the error mean under control from OPEN to PREMIER for illustrative purposes.

We fit a linear model to determine which covariates significantly impact the measurement error under control conditions, using data from both PREMIER and OPEN (see Supplemental Table \ref{tab:regoutput}). The error under control appears to be differential with respect to sex and race, and also weakly differential with respect to education. Given the output of this model, and the covariate distributions shown in Table \ref{tab:covtab}, we therefore have reason to believe that the transportability assumption is violated.

Next, we fit a sample membership model using all five covariates, which includes covariates that impact both the measurement error and sample membership, as well as covariates that just differ between the two samples. To fit the model, we use generalized boosted models (GBM), an algorithm that allows for flexible, nonlinear relationships when modeling study membership by a large number of covariates \citep{mccaffrey2004propensity}. We examine the distributions of predicted sample membership probabilities (see Supplemental Figure \ref{fig:ps}), which have an $ASMD = 1.47$. This is unsurprising, given the large differences between the two samples' distributions of race and BMI. There are some outliers in the resulting validation sample weights that are more than ten times the average of the weights. We therefore implement weight trimming to account for the extreme weights, setting all validation sample weights in the top decile to the 90th percentile weight value \citep{lee2011weight} (see Supplemental Figure \ref{fig:weight_hist} for the distribution of the trimmed weights in the validation sample).

Lastly, we use the weights to estimate $\hat\mu_0^{weighted}$. Table \ref{tab:dataresult} shows both the unweighted and weighted estimates, along with the estimate of $\mu_0^{rct}$ in PREMIER (which again, is usually not estimable when only self-reported outcomes are collected). Observe that for the outcome at 18 months, the absolute bias of the $\hat\mu_0$ estimate, $\text{bias}_{\hat\mu_0}$, decreases by about 80\% after implementing the weighting method.

\begin{table}

\caption{Estimated Error Mean under Control Conditions in the Validation Sample, and Associated Bias, by Weighting Method \label{tab:dataresult}}
\centering
\begin{tabular}[t]{ccccccc}
\toprule
\multicolumn{1}{c}{} & \multicolumn{3}{c}{6 months} & \multicolumn{3}{c}{18 months} \\
\cmidrule(l{3pt}r{3pt}){2-4} \cmidrule(l{3pt}r{3pt}){5-7}
Method & $\hat\mu_0$ & $\mu_0^{rct}$ & $\text{bias}_{\hat\mu_0}$ & $\hat\mu_0$ & $\mu_0^{rct}$ & $\text{bias}_{\hat\mu_0}$\\
\midrule
Unweighted & -0.228 &  & -0.001 & -0.228 &  & 0.077\\
\cmidrule{1-2}
\cmidrule{4-5}
\cmidrule{7-7}
Weighted & -0.326 & \multirow{-2}{*}{\centering\arraybackslash -0.227} & -0.099 & -0.321 & \multirow{-2}{*}{\centering\arraybackslash -0.305} & -0.016\\
\bottomrule
\end{tabular}
\end{table}

In the data example, note that the ASMD of the sample membership probabilities between OPEN and PREMIER is quite large (1.47), and that some of the covariates are extremely different from one another. Even after weighting, OPEN is still a bit dissimilar from PREMIER by BMI and race (see Supplemental Figure \ref{fig:loveplot}). Additionally, by fitting the selection model using GBM, we are able fitting a model somewhere between the true form and the main-effects-only form. We therefore see that these results reflect the simulation findings under rather extreme cases, suggesting that the weighting may help to a certain extent, but that the differences between the validation sample and trial may lend to sub-optimal performance. 

\section{Discussion}
\label{s:discuss}

When using self-reported measures as outcomes in a lifestyle intervention study, it is important to correct for any potential measurement error in order to make accurate inferences on the effect of the treatment in the study population. While measurement error is a well documented issue, particularly in nutritional epidemiology, there is still much need for increased method implementation in applied research studies, as well as improved methodology for different types of error \citep{jurek2006exposure, brakenhoff2018measurement}. We highlight the importance of considering transportability when utilizing external validation studies to correct for outcome measurement error. Using externally estimated measurement error may introduce additional biases to the ATE estimate in cases where validation samples are dissimilar from the primary intervention study of interest. We show that weighting the validation sample to better resemble the intervention study can reduce such biases, and improve upon the transportability of the measurement error estimated in the external sample. However, in some extreme cases, it may still be inappropriate to transport if the validation sample is vastly different from the trial on a set of observed characteristics. Additionally, it is important to remember that while researchers are often concerned about measurement error, it will only lead to a biased ATE estimate when the outcome error is differential with respect to treatment and the error means are thereby different across treatment groups. Such bias would prompt the usage of external validation data for outcome measurement error correction, and thus the concerns about transportability (see Table \ref{tab:biascases}).

There are several limitations to the work presented in this paper. First, we assume that the measurement error model structure (i.e. the model relating the measurement error to the covariates) in both the intervention study and the validation sample are the same. It is possible that such relationships may differ between studies, even though in practice, this would be untestable without observing the outcome without measurement error in the trial itself. Further research is needed to understand transportability and to apply the methods proposed in this paper when relaxing this assumption. Second, we assume in this work that we are able to fully observe all covariates that impact the outcome measurement error structure in both the intervention study and the validation sample. Due to data availability, certain important variables may be unobserved in practice, either in one of the datasets, or in both, which may hinder the performance of these methods. Sensitivity analyses should be adapted and applied to address these concerns \citep{nguyen2017sensitivity}. Lastly, as seen with PREMIER, transportability may vary by time-point with longitudinal outcomes, warranting further investigation into how transportability and measurement error may vary over time.

This work has focused on the use of external validation samples only, and further research is needed to evaluate the differences in transportability between internal and external validation samples. In some cases, internal validation samples may still be preferable when possible to incorporate into study design, particularly such that true outcome measures can be obtained under different treatment conditions. When it is infeasible to collect internal validation data, researchers designing external validation studies should still consider the possible relevant trial study populations to which the validation sample will be applied to. Taking such steps when designing validation studies will also help ensure better transportability when using information from external data sources to correct for outcome measurement error.



\section*{Supplemental Material}
\subsection*{R Code}
All code for the simulation and data example can be found in the following GitHub repository:

https://github.com/benjamin-ackerman/ME\_transportability 

\subsection*{Derivation of Standard Error for $\hat\mu_0^{weighted}$}
Recall that 

\begin{equation*}
\hat\mu_0^{weighted} = \frac{\sum_{i=1}^{n_v} w_i(Y_i-Z_i)}{\sum_{i=1}^{n_v}w_i}
\end{equation*}

Let $Y^*$ and $Z^*$ be the weighted vectors of $Y$ and $Z$ in the validation sample. Since $Y$ and $Z$ are paired measurements and are not independent, $\text{var}(Y^* - Z^*) = \text{var}(Y^*) + \text{var}(Z^*) - 2\text{cov}(Y^*,Z^*)$.

According to \citet{price1972extension}, $$\text{cov}(Y^*, Z^*) = \frac{\big[ \sum_{i}^{n_v} w_i (Y_i - \bar Y^*)(Z_i - \bar Z^*)\big]}{\sum_{i}^{n_v} w_i} $$

where 
\begin{align*}
\bar Y^* &= \frac{\sum_{i=1}^{n_v} w_i Y_i}{\sum_{i=1}^{n_v}w_i}\\
\bar Z^* &= \frac{\sum_{i=1}^{n_v} w_i Z_i}{\sum_{i=1}^{n_v}w_i}\\
\text{var}(Y^*) &= \text{cov}(Y^*, Y^*)\\
\text{var}(Z^*) &= \text{cov}(Z^*, Z^*)
\end{align*}

Therefore, $\text{se}_{\hat \mu_0^{weighted}} = \sqrt{\text{var}(Y^* - Z^*)/n}$ and can be used to construct a 95\% confidence interval.

\subsection*{Supplemental Tables and Figures}
\begin{table}[!h]
\centering
\caption{Scaled coefficients for covariates by model type}\label{tab:simparams}
\begin{tabular}{l|llll|}
\cline{2-5}
        & $X_1$ & $X_2$ & $X_3$     & $X_4$     \\ \hline
\multicolumn{1}{|l|}{S model} & $\gamma_1$ & 0    & $\frac{1}{2}\gamma_1$ & $2\gamma_1$   \\ \hline
\multicolumn{1}{|l|}{Y model} & 0    & $\gamma_2$ & $2\gamma_2$   & $\frac{1}{2}\gamma_2$ \\ \hline
\end{tabular}
\end{table}

\begin{table}[!h]
\caption{Self-reported and Urinary Sodium Outcomes by Study and Treatment Group \label{tab:outcomes}}
\centering
\begin{tabular}{lrrrrr}
\toprule
\multicolumn{1}{c}{} & \multicolumn{1}{c}{OPEN} & \multicolumn{4}{c}{PREMIER} \\
\cmidrule(l{3pt}r{3pt}){2-2} \cmidrule(l{3pt}r{3pt}){3-6}
\multicolumn{1}{c}{} & \multicolumn{1}{c}{} & \multicolumn{2}{c}{6 months} & \multicolumn{2}{c}{18 months} \\
\cmidrule(l{3pt}r{3pt}){3-4} \cmidrule(l{3pt}r{3pt}){5-6}
  & Control & Control & Treatment & Control & Treatment\\
\midrule
Self-Reported (Y) & 8.220 & 7.850 & 7.640 & 7.840 & 7.690\\
Urine (Z) & 8.450 & 8.070 & 7.970 & 8.140 & 8.020\\
$\mu_{a}^{s}$ & -0.227 & -0.228 & -0.323 & -0.299 & -0.327\\
\bottomrule
\end{tabular}
\end{table}

\begin{table}[!h]

\caption{T-test comparing measurement error by treatment group in PREMIER \label{tab:errortx}}
\centering
\begin{tabular}{lrlr}
\toprule
Timepoint & $\text{bias}_{\hat \Delta}$ & $95\%$ CI & p-value\\
\midrule
6 months & 0.1108736 & (0.0162, 0.206) & 0.0217971\\
18 months & 0.0282937 & (-0.065, 0.122) & 0.5517550\\
\bottomrule
\end{tabular}
\end{table}

\begin{table}[!h]

\caption{Regression Coefficients for modeling effect of covariates on measurement error term \label{tab:regoutput}}
\centering
\begin{tabular}{lrrrr}
\toprule
term & estimate & std.error & statistic & p.value\\
\midrule
Intercept & 5.90 & 0.41 & 14.53 & 0.0000000\\
log(sodium urine) & 0.24 & 0.05 & 4.74 & 0.0000029\\
Sex & 0.20 & 0.04 & 5.33 & 0.0000002\\
Age 41-45 & 0.06 & 0.10 & 0.65 & 0.5178708\\
Age 46-50 & 0.06 & 0.10 & 0.63 & 0.5302275\\
\addlinespace
Age 51-55 & -0.07 & 0.10 & -0.78 & 0.4365246\\
Age 56-60 & -0.06 & 0.10 & -0.61 & 0.5419625\\
Age > 60 & 0.00 & 0.09 & 0.03 & 0.9789308\\
BMI & 0.00 & 0.00 & 1.21 & 0.2275847\\
College Education & 0.09 & 0.05 & 1.72 & 0.0864365\\
\addlinespace
Grad School Education & 0.10 & 0.05 & 1.84 & 0.0668818\\
Black & -0.28 & 0.07 & -3.89 & 0.0001163\\
\bottomrule
\end{tabular}
\end{table}

\begin{figure}
 \centerline{\includegraphics[width=145mm]{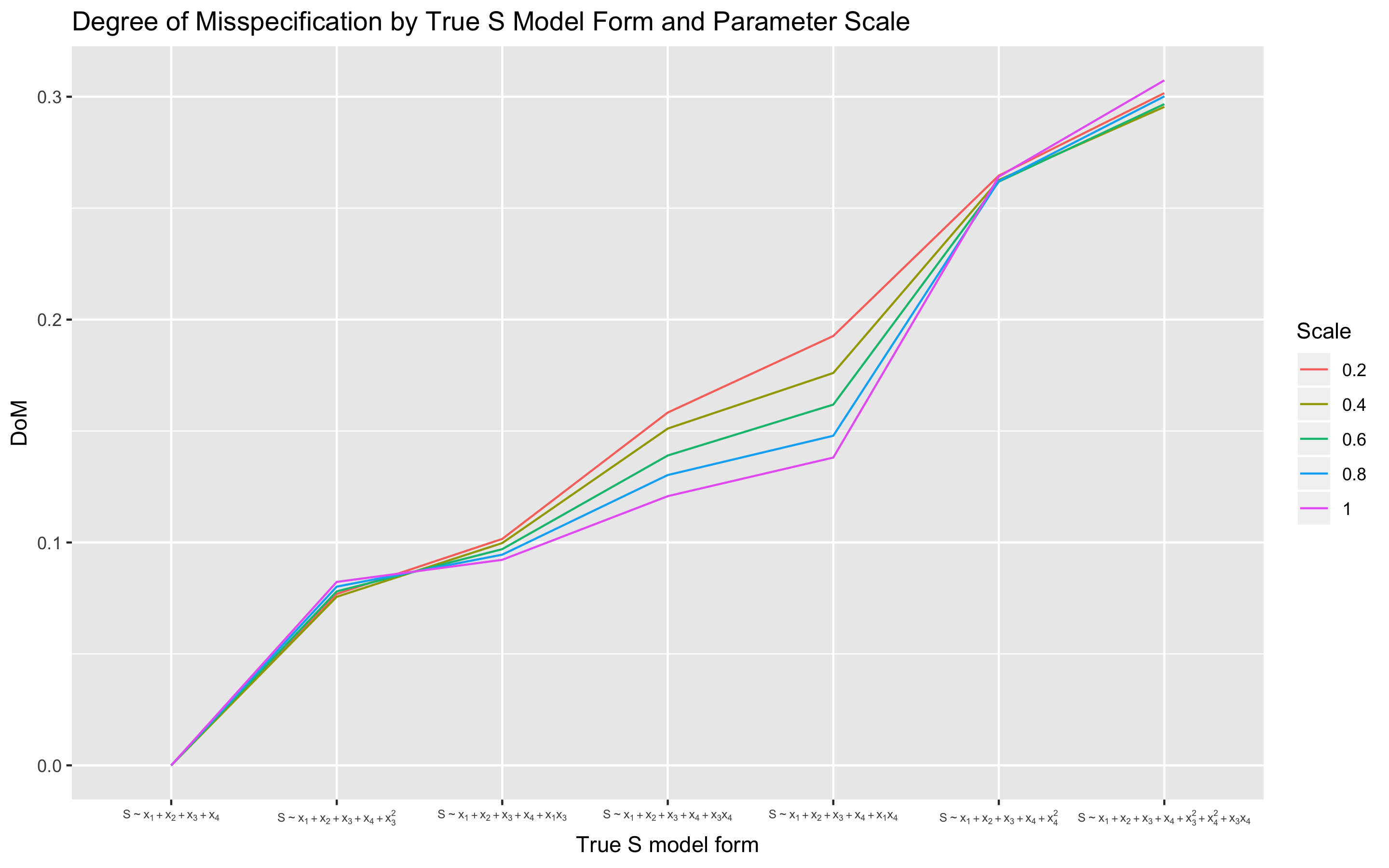}}
\caption[Degree of Misspecification (DoM) of fitting the main effects model under different true S model forms]%
{Degree of Misspecification (DoM) of fitting the main effects model under different true S model forms. As expected, when the true model is the main effects model, there is no misspecification, and the most misspecified model is the model with the most terms. Also note that the degree of misspecification increases when the true S model form depends on the more influential covariates.}
\label{fig:dommodel}
\end{figure}

\begin{figure}
 \centerline{\includegraphics[width=145mm]{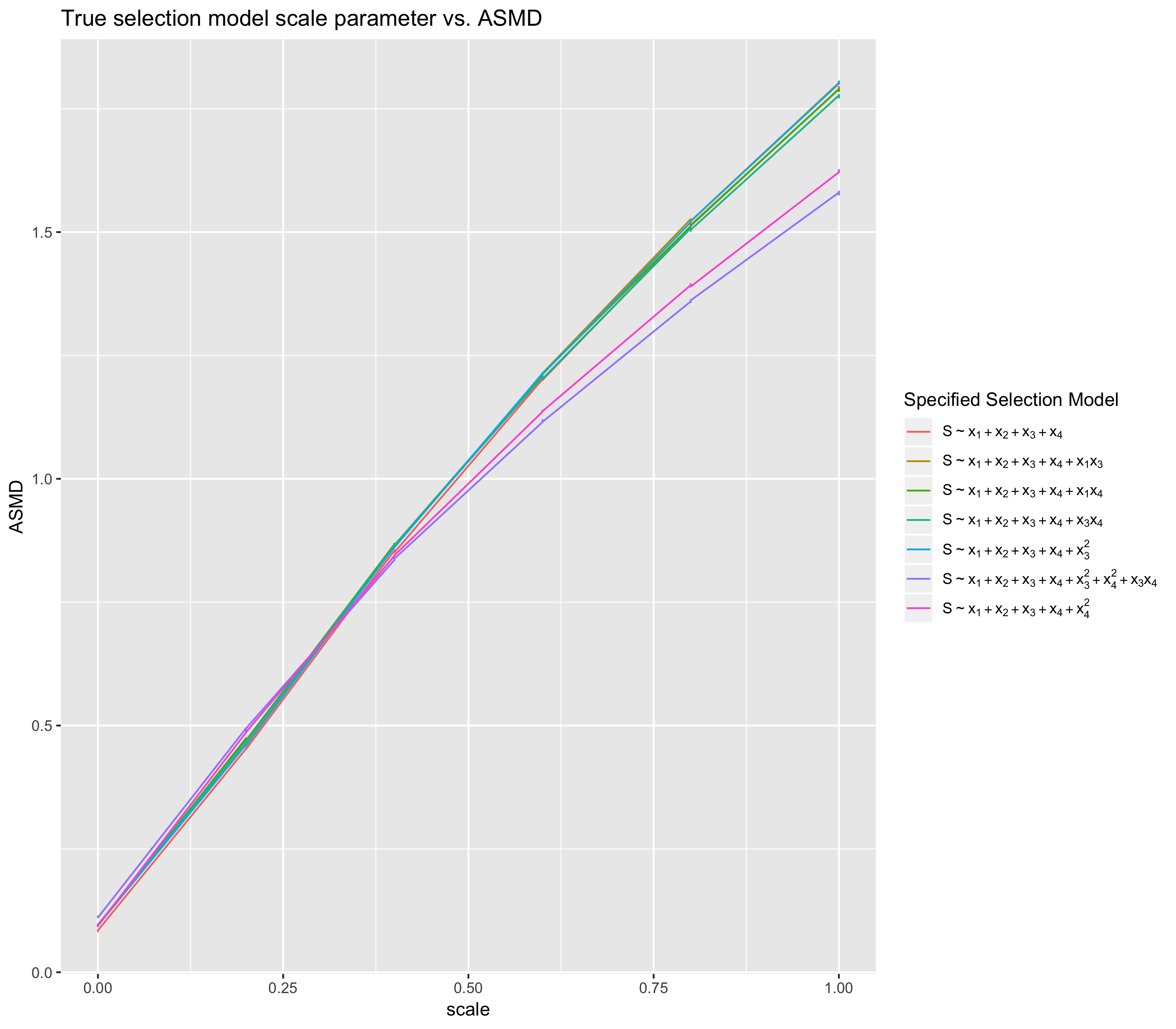}}
 \caption[The relationship between the `scale' parameter ($\gamma_1$) and the ASMD between the true selection probabilities across the trial and the validation data]%
 {The relationship between the `scale' parameter ($\gamma_1$) and the ASMD between the true selection probabilities across the trial and the validation data. There's slight variation across the different true selection models, though this could be due to simulation variability.} \label{fig:asmdscale}
\end{figure}

\begin{figure}
 \centerline{\includegraphics[width=145mm]{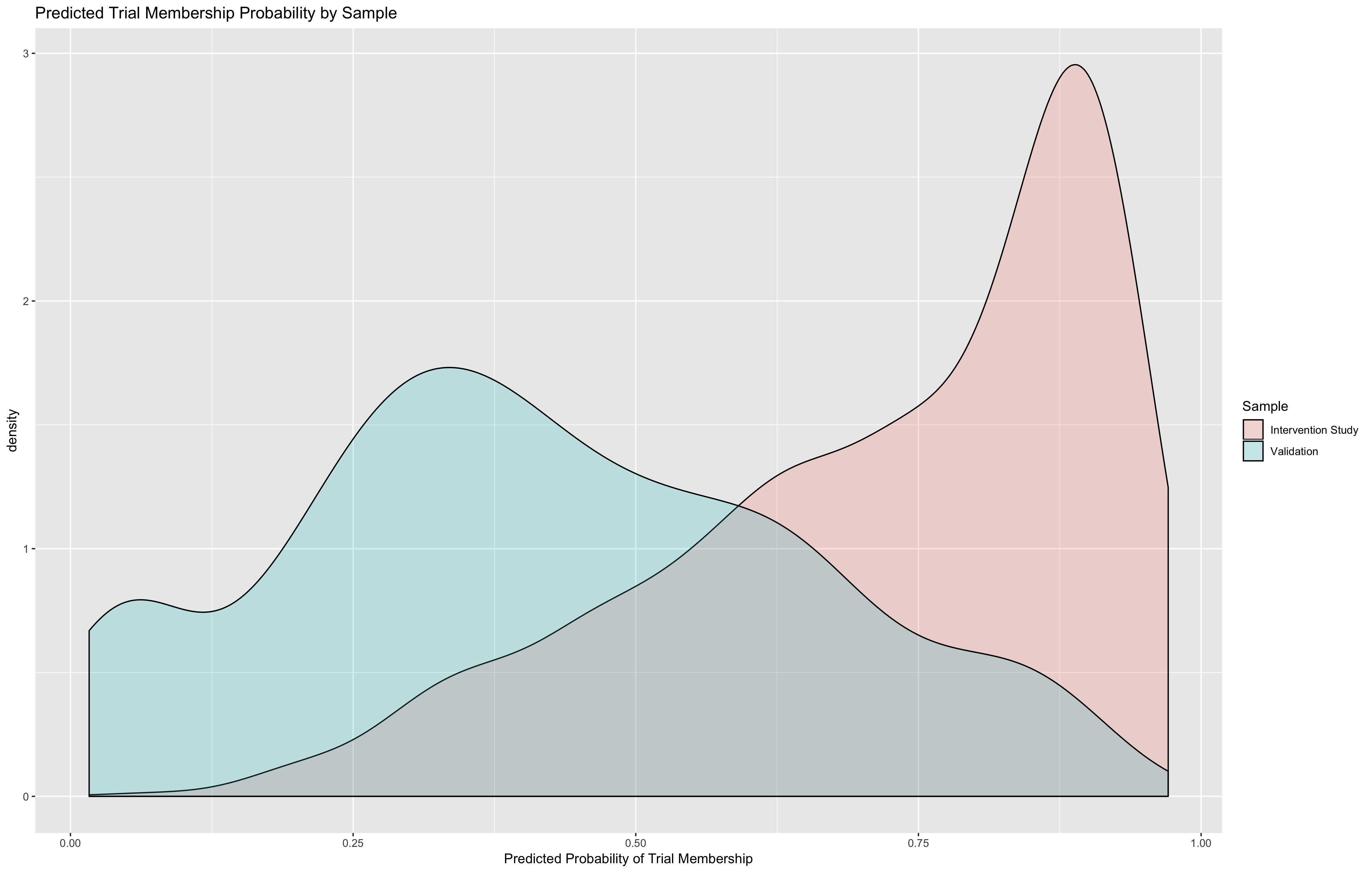}}
\caption{Predicted Probabilities of Trial Membership by Sample} \label{fig:ps}
\end{figure}

\begin{figure}
 \centerline{\includegraphics[width=145mm]{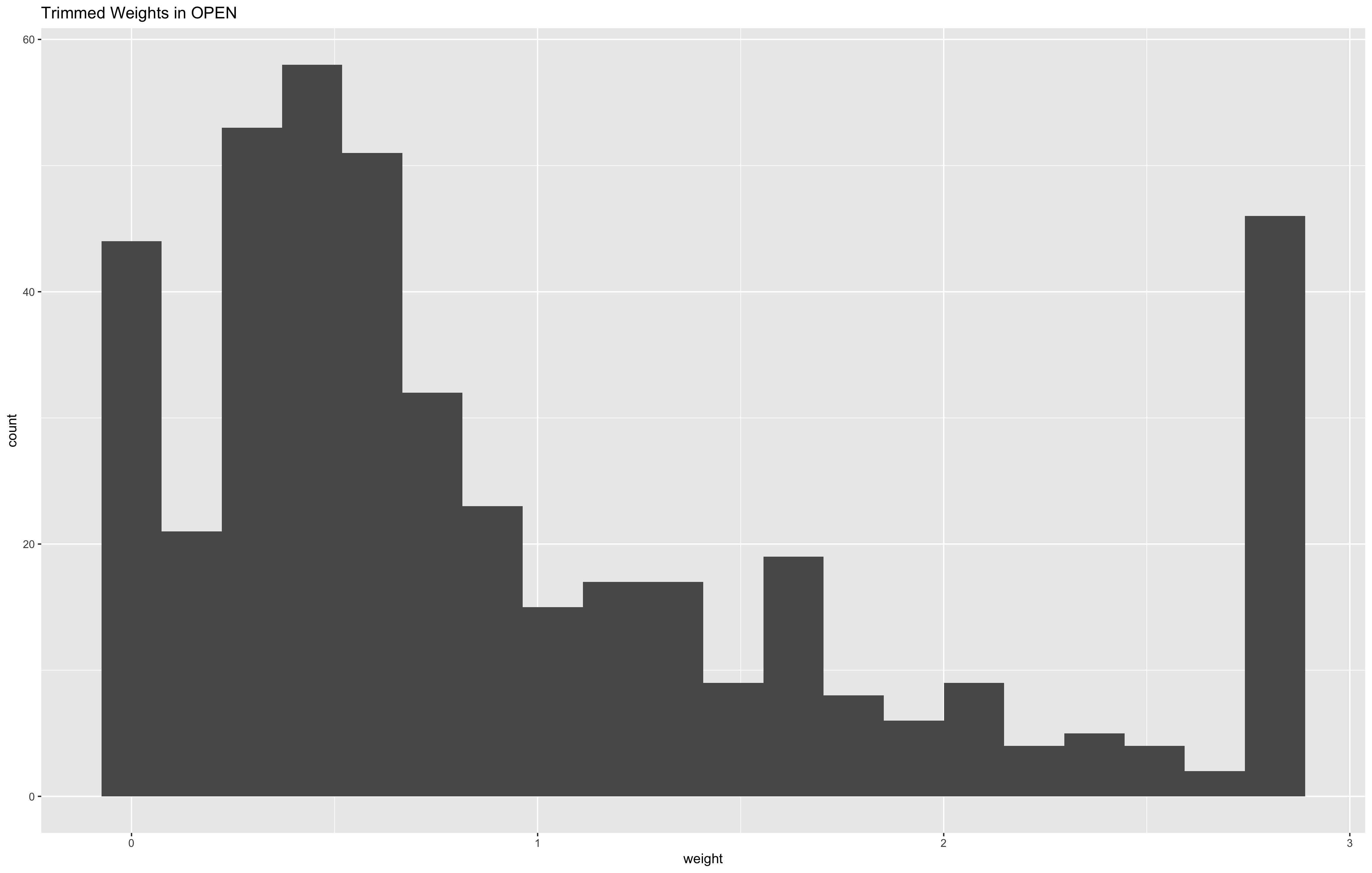}}
\caption{Distribution of the trimmed weights in OPEN validation sample} \label{fig:weight_hist}
\end{figure}

\begin{figure}
 \centerline{\includegraphics[width=145mm]{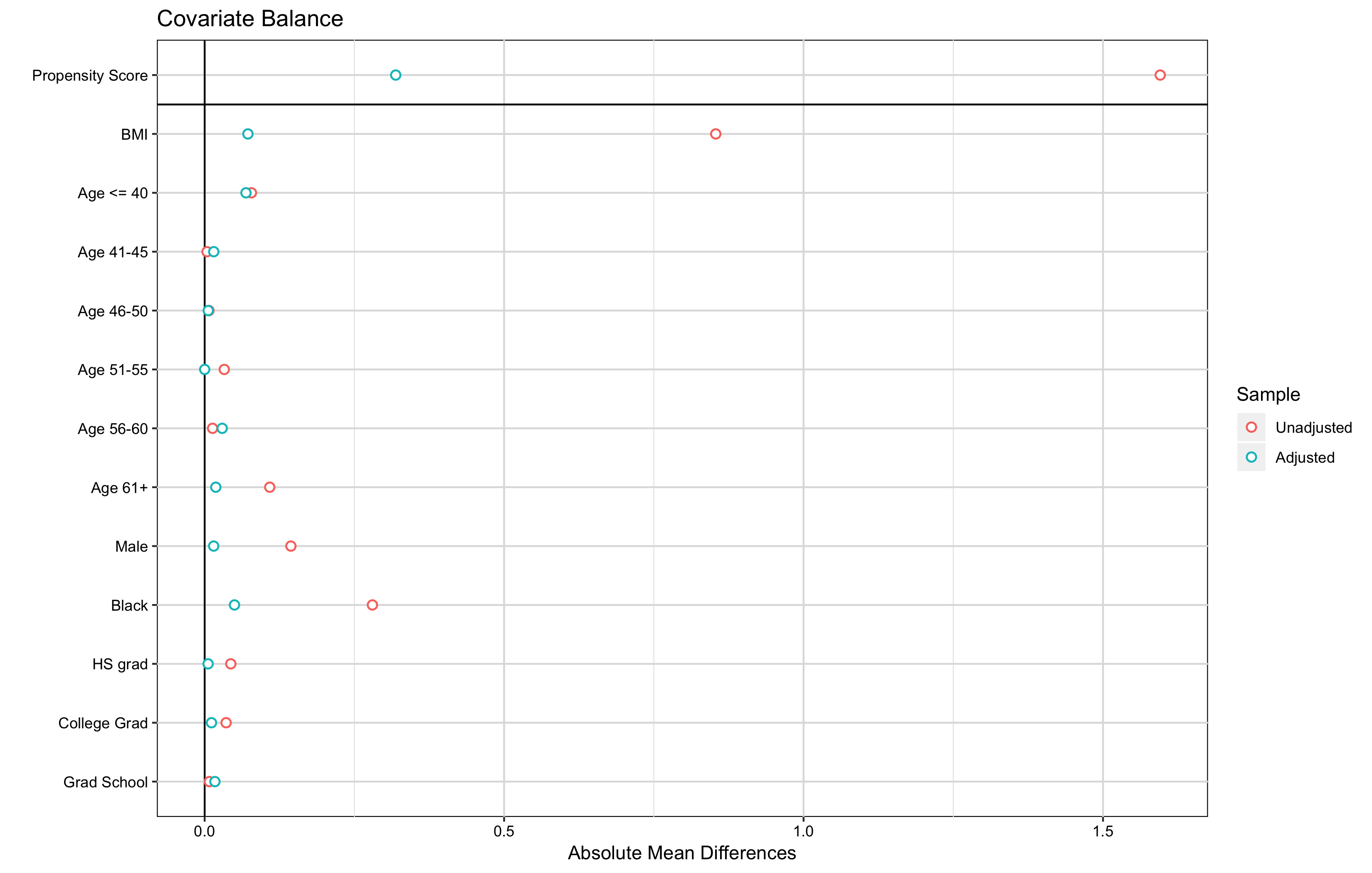}}
\caption{Love plot comparing the covariate distributions in the intervention trial PREMIER (pink) and validation sample OPEN (blue), pre-and-post weighting the validation sample} \label{fig:loveplot}
\end{figure}

\end{document}